\documentclass[12pt]{article}
\usepackage[usenames,dvipsnames,svgnames,table,x11names]{xcolor}
\usepackage{kpfonts}
\usepackage{packages-csm}
\usepackage{simeontex-csm}
\usepackage{definitions-csm}
\usepackage[utf8]{inputenc} 
\usepackage{tikz}
\usetikzlibrary{arrows,decorations.markings}
\usetikzlibrary{calc}
\usepackage{slashed}
\usepackage{braket}
\usepackage[affil-it]{authblk}
\usepackage[outdir=../fig/]{epstopdf}

\definecolor{refkey}{rgb}{0.85, 0.0, 0.3}
\definecolor{labelkey}{rgb}{0.2,0.2,0.6}

\preprint{{IPMU17-0081}\\{CALT-TH-2017-24}}

\title{A Note on Inhomogeneous Ground States at Large Global Charge}

\author[1]{Simeon Hellerman}
\author[1,2]{Nozomu Kobayashi}
\author[1,2]{Shunsuke Maeda}
\author[1,2]{Masataka Watanabe}
\affil[1]{\small Kavli Institute for the Physics and Mathematics of the Universe (WPI),The University of Tokyo Institutes for Advanced Study, The University of Tokyo, Kashiwa, Chiba 277-8583, Japan}
\affil[2]{\small Department of Physics, Faculty of Science, The University of Tokyo, Bunkyo-ku, Tokyo 133-0022, Japan}
\date{}

\allowdisplaybreaks[4]

\def\ellz{\ell}
\begin{document}

\hypersetup{linkcolor=black}
  \maketitle 
     \abstract{
In this note we search for the ground state, in infinite volume, of the $D=3$
Wilson-Fisher conformal $O(4)$ model, at nonzero values
of the two independent charge densities $\rho_{1,2}$.  Using
an effective theory valid on scales longer than the scale
defined by the charge density, we
show that the ground-state configuration is inhomogeneous for generic ratios
$\rho_1 / \rho_2$. This result confirms, within the context of a well-defined effective
theory, a recent no-go result of \cite{Alvarez-Gaume:2016vff} .   We also show
that any spatially
periodic ground state solutions have an energetic preference
towards longer periods, within some range of $\rho_1 / \rho_2$ containing
a neighborhood of zero.  This suggests that the scale of variation
of the ground state solution in finite volume will be the infrared scale,
and that the use of the effective theory at large charge in finite volume
is self-consistent.}
      \newpage
\setcounter{tocdepth}{3}
\setcounter{secnumdepth}{4}
\tableofcontents
\hypersetup{linkcolor=PaleGreen4}
\newpage

\section{Introduction}

Conformal field theories with global symmetries display interesting and useful
simplifications in the sector of large global charge.  These simplifications
make it possible to calculate asymptotic expansions of charged
operator dimensions and OPE coefficients 
\cite{Hellerman:2015nra,Alvarez-Gaume:2016vff,Monin:2016jmo} to any desired accuracy in terms of a small number of undetermined
coefficients in an effective Lagrangian describing the local dynamics of
the system at large charge density.  Although these calculations use
lagrangian methods, the results are strikingly parallel to the large-spin
expansion of operator dimensions obtained by the light-cone bootstrap
\cite{Komargodski:2012ek, Fitzpatrick:2012yx}
and along with those results, work best in a regime of large charge
and large operator dimensions,
complementing the regime of $O(1)$ charges and operator dimensions
 \cite{Kos:2016ysd, Kos:2015mba, Kos:2013tga, Dey:2016zbg, Diab:2016spb}
that is accessed efficiently by numerical linear programming methods of solving the conformal bootstrap \cite{Rattazzi:2008pe , ElShowk:2012ht , El-Showk:2014dwa}.

In
\cite{Hellerman:2015nra,Alvarez-Gaume:2016vff,Monin:2016jmo},
the properties of charged local operators are calculated in radial quantization by
quantizing the large-charge effective Lagrangian on a spherical
spatial slice.  The
hierarchy between the ultraviolet scale $E\ll{\rm UV} \equiv \r\uu{{1\over{D-1}}}$
and the infrared scale, $E\ll{\rm IR} = r\uu{-1}\ll{\rm sphere}$ is
$E\ll{\rm IR} / E\ll{\rm UV} \propto J\uu{-{1\over{D-1}}}$, where
$J$ is the global charge of the local operator in the CFT.
This large hierarchy renders
the large-charge effective Lagrangian weakly coupled, and allows
the perturbative computation of CFT data with quantum corrections and higher-derivative operators in the large-charge EFT, suppressed by inverse powers
of $J$.

In order to get started on such a calculation, one needs to know the structure
of the large-charge effective lagrangian, and the nature of the ground state
carrying a given set of global charges.  In the limit where the charge is
taken to infinity, one can try to flatten out the sphere and consider the
system in infinite flat space at fixed charge density $\r$.  Naively, then,
it would seem that each large-charge limit in a CFT should correspond
to a homogeneous ground state of a CFT with a chemical potential.  And 
indeed, various interesting new phases of matter with spontaneously broken
conformal and Lorentz symmetries have been derived through these
considerations 
\cite{Hellerman:2015nra,Alvarez-Gaume:2016vff,Monin:2016jmo}.

The expectation that the large-charge limit always defines a homogeneous
phase of matter is a bit too naive however, as it assumes the classical solution
describing the large-charge ground state on
the sphere, is spherically symmetric.  It is
interesting to note that this expectation can be proven false in some very simple
cases.  In 
\cite{Alvarez-Gaume:2016vff}, the authors studied the conformal Wilson-Fisher
$O(2N)$ model \cite{Wilson:1971dc} in $D=3$
at large Noether charge, and found that 
a homogeneous ground state in flat space exists only in the case where the element of
the adjoint of $O(2N)$ defined by the total charge, 
has minimal rank, which is to say a single nonvanishing antisymmetric
$2\times 2$ block , and zeroes everywhere else.  In the case
where the charge matrix has minimal nonzero rank, the homogenous ground
state in flat space was studied in detail and many interesting properties
extracted.  Left unanswered is the question of the nature
of the ground state when the charge matrix has nonminimal rank.

In this note, we will address this question in the simplest nontrivial example,
that of
the Wilson-Fisher $O(4)$ fixed point in three dimensions.  We will 
find that there are no exactly homogeneous ground states, 
but a family of inhomogeneous, spatially periodic
solutions of arbitrarily large spatial
period in infinite volume, with an energetic preference for
longer spatial periods in some range of ratios of the two independent
charges.  In this range of charges, then, the system will be driven dynamically
into a regime where the fields vary slowly on the scale of the
charge density, and the large-charge effective
Lagrangian is parametrically reliable.

\section{The $O(4)$ model at finite charge density}

We now analyze the $O(4)$ model in infinite volume, for general global
charges.  That is, we examine what
the ground state looks like when we let the charge be proportional
to a general element of the adjoint of $SO(4)$.  We first refine and make more rigorous the result
of 
\cite{Alvarez-Gaume:2016vff} by following the recipe
of 
\cite{Hellerman:2015nra}, integrating
out the heavy mode and working strictly within a conformal sigma model
that is singular in the vacuum but nonsingular around a state with large charge
density.

In this framework, we rigorously
reproduce the no-go result of 
\cite{Alvarez-Gaume:2016vff}: We find that all candidate ground state solutions are inhomogeneous,
and break the translational symmetry in one direction down to at most a discrete
subgroup with period $\ell$, if the antisymmetric matrix defining the charge
has nonvanishing determinant.   In particular, for each value of
the ratio $\hat{\r}\ll 1 / \hat{\r}\ll 2$ of the two eigenvalues of the charge matrix, there
is one spatially periodic solution with period $\ell$ that also satisfies
a helical symmetry, \it i.e., \rm a symmetry under combined time translation
and global symmetry rotation.

This raises two closely related questions:  First, which candidate periodic 
solution is
the true ground state?  That is, which value of the spatial period $\ell$, if any,
minimizes the energy for given global charge densities?  Second,
for what range of $\ell$ is the effective field theory reliable?

\subsection{Parametrizing the charge density}

To answer these questions quantitatively, we must
find a convenient way to express the charge density itself, as an element
of the adjoint of $SO(4)$, that is, a general
$4\times 4$ imaginary antisymmetric matrix.  Such a matrix has
real eigenvalues that occur in pairs with equal magnitude
and opposite sign.  The two independent positive eigenvalues are $\hat{\r}\ll{1,2}$

Rather than parametrizing the charge density directly by the two
independent eigenvalues
$\hat{\r}\ll{1,2}$ of the charge matrix, we follow 
\cite{Alvarez-Gaume:2016vff} in
choosing a basis for the  chemical potential, which is equivalent to
diagonalizing the generator defining the symmetry of the helical solution.
Choose a complex basis for the fundamental of
$U(2) \subset SO(4)$, and parametrize the charge generator by
the two matrix elements $\r\ll{1,2}$ on the diagonal.  This will turn
out to be equivalent: For helical solutions, the charge matrix commutes with
the chemical potential, its off-diagonal terms
always vanish, and $\hat{\r}\ll 1 / \hat{\r}\ll 2$ is simply equal to
$\r\ll 1 / \r\ll 2$.  

We will see that there is an unstable direction of the classical solution, such that minimizing
the energy at fixed charge densities in infinite volume, leads to an instability
towards an infinite spatial period, for sufficiently small values
of the ratio $\r\ll 1 / \r\ll 2$.
  
For purposes of computing the operator spectrum in radial
quantization, we would ultimately
want to put the theory on $\mathbb{R}\times S^2$, but in the present note
we will aim to understand some local aspects of the charged ground state by
taking a limit of large charge and fixed average density, which amounts to
quantizing the theory on $\mathbb{R}\times \mathbb{R}^2$.  We will comment
in the Discussion section on the relevance to the ground state in finite volume.

\subsection{Conformal sigma model from linear 
sigma model}

The O(4) model is described by four real scalars $X\uu{1,2,3,4}$, which
we organize into a complex $SU(2)$ doublet $Q\equiv
 \begin{pmatrix}
X\ll 1 + i X\ll 2 \\ X\ll 3 + i X\ll 4
\end{pmatrix}$.
The $O(4)$ critical point is obtained by starting in the ultraviolet, giving the scalars
a quartic potential proportional to $(X\sqd)\sqd = |q|\uu 4,$ and fine-tuning the 
mass term $m\sqd |Q|\sqd$ to the unique strength such that the system has infinite correlation length
and flows to a nontrivial fixed point of the renormalization group.
   
We wish to parametrise $Q$ as follows in terms of amplitudes and angles:
\begin{equation}
Q=A\times q,\quad q=
\begin{pmatrix}
q_1\\ q_2
\end{pmatrix},
\end{equation}
where $|q_1|^2+|q_2|^2=1$.
We can expand the solution
at large $A$.  The leading action at large and approximately
constant $A$ is sextic potential which is generated along
the RG flow as explained in 
\cite{Hellerman:2015nra}.

The Lagrangian of the theory in the IR becomes
\begin{equation}
\mathcal{L}_{\mathrm{IR}}=\frac{1}{2}(\partial A)^2+\frac{\gamma}{2}A^2\partial q^\dagger \partial q-\frac{h^2}{6}A^6,
\label{IRLagrangian0}
\end{equation}
under a field-reparametrization condition that the kinetic term of $A$ is canonical. 

We have omitted other terms as well.  In the present note we use only the leading large-density term, and so we omit the Ricci coupling and higher derivative terms.  
The justification for the omission of these terms, is important and we
must consider it carefully.  Higher-derivative terms are suppressed
when the fields vary on scales $L$ which are long compared
to the ultraviolet scale $(\r\ll 1 + \r\ll 2)\uu{-\hh}$. For values of $L$
smaller than $(\r\ll 1 + \r\ll 2)\uu{-\hh}$, the large-charge effective
theory is not within its range of validity, because the conformal goldstone
fields are varying rapidly on the scale of the charge density itself.  Higher-derivative operators and quantum corrections are unsuppressed, and there
is no obvious simplification of the dynamics.

For generic charge densities $\r\ll{1,2}$, we will find that there is no
homogeneous ground state classical solution, so the question of the scale of variation $L$ of the classical solution is crucial.  If the ground state
of the system has $L$ smaller than or comparable to $(\r\ll 1 + \r\ll 2)\uu{-\hh}$,
then the use of the effective theory is not allowed.  If the ground
state of the system only has inhomogeneities on scales $L$ 
longer than $(\r\ll 1 + \r\ll 2)\uu{-\hh}$, the use of the EFT will be justified.
We will see that the latter situation holds for some range of $\r\ll 1 / \r\ll 2$
that contains a neighborhood of zero.  
For now, simply assume the fields are slowly varying
on the scale set by the density itself and the effective theory will
be usable; we will then justify this assumption \it a posteriori. \rm

With this assumption, the field $A$ has a mass scale
set by the density, and therefore should be integrated out in such a limit.
The equilibrium value of $A$ is given by 
\begin{equation}
\frac{\delta \mathcal{L}_{\rm IR}}{\delta A}=0
\iff A^2=\sqrt{\frac{\partial q^\dagger \partial q}{\gamma^{-1} h^2}}
\end{equation} 
Plugging this into \eqref{IRLagrangian0}, we get the conformal sigma model on $S^3$ as follows:
\begin{equation}
\mathcal{L}=b_q\mathcal{L}_0^{3/2}=b_q(\partial q^\dagger \partial q)^{3/2},
\label{conformalsigma}
\end{equation}
where $|q| = 1$ and $b_q=\sqrt{\gamma^3h^{-2}}/3$ is an undetermined coefficient which should come from the complicated, original RG flow equation,
as in \cite{Hellerman:2015nra}.

\subsection{Restriction to fixed average charge densities $\r\ll{1,2}$}
Because we are putting the theory on $\mathbb{R}^2$, and the concept of total charge is ill-defined, we can only fix the average charge density instead of the total charge itself.  We impose the following conditions unto Noether currents: 
\begin{eqnarray}
-\left.\frac{2ib_q}{3}\int dx^{i} \sqrt{\mathcal{L}_0}\left[q^{\dagger}\partial_t q {{{}-\mathrm{{}c.c.}}}\right]\right/\mathcal{V}&=&\rho_1+\rho_2\label{charge}
\\
-\left.\frac{2ib_q}{3} \int dx^{i} \sqrt{\mathcal{L}_0}\left[q^{\dagger}\sigma^{3}\partial_t q{{{}-\mathrm{{}c.c.}}}\right]\right/\mathcal{V}&=&\rho_1-\rho_2, 
\label{singlet}
\end{eqnarray}
where $\mathcal{V}$ indicates the total volume of the space.
Under these constraints, we look for a field configuration that has
the lowest energy, whose density is given by
\begin{equation}
\mathcal{H}=
 b_q\sqrt{\dot{q}^\dagger\dot{q}-\partial_i q^\dagger \partial^i q}\times \left(2\dot{q}^\dagger\dot{q}+\partial_i q^\dagger \partial^i q\right)
 \label{energydensity}
\end{equation}

\subsection{Equation of motion for the conformal sigma model}
Now we are ready to derive the equation of motion for \eqref{conformalsigma}.
We set an ansatz for the ground state solution that it is at least homogeneous in the one of the spatial directions, the $y$ direction, and varies spatially only
in the $x$ direction.

We also use the fact that the time dependence of
the ground state solution must be helical, and also that it is invariant under the combination of $t\to -t$ and complex conjugation.  Then the ground state solution for $q$ can be parametrised as follows:
\begin{equation}
q=
\begin{pmatrix}
q_1\\ q_2
\end{pmatrix}
=
\begin{pmatrix}
e^{i\omega_1 t}\sin(p(x))\\ e^{i\omega_2 t}\cos(p(x))
\end{pmatrix},
\end{equation}
where we are free to set $\omega_1>\omega_2$
The equation of motion for the $p$ field is then
\begin{equation}
\mathcal{L}-p^{\prime}(x)\frac{\delta \mathcal{L}}{\delta p^{\prime}}
=\rm{(const.)}
\end{equation}
Using a constant $\kappa$ that is of the same mass dimension as $\omega_{1,2}$, we rewrite the above equation as
\begin{equation}
-\frac{\kappa^6}{4}=-\frac{b_q^{-2}T_{xx}^2}{4}=\left(p^{\prime}(x)^2-V(p(x)))\right)
\left(p^{\prime}(x)^2+\frac{V(p(x))}{2}\right)^2,
\end{equation}
where 
\begin{equation}
V(p)=\omega_2^2+(\omega_1^2-\omega_2^2)\sin^2(p).
\end{equation}

The constraints imposed by \eqref{charge} and \eqref{singlet} become
\begin{eqnarray}
\rho_1&=&\frac{8b_q}{3\mathcal{V}}\int dx^i\, \omega_1\sqrt{-p^{\prime}(x)^2+V(p(x)))}\sin^2(p(x))\label{baryonsin}\\
\rho_2&=&\frac{8b_q}{3\mathcal{V}}\int dx^i\, \omega_2\sqrt{-p^{\prime}(x)^2+V(p(x)))}\cos^2(p(x)),\label{baryoncos}
\end{eqnarray}
Notice from the equation of motion that the solution for $p$ is inevitably inhomogeneous unless $\omega_1=\omega_2$, which will never be the case if both $\rho_1$ and $\rho_2$ are nonvanishing.
This means that the charged ground state configuration for the $O(4)$ theory is generically inhomogeneous, as promised in the introduction and demonstrated in the context of the model of 
\cite{Alvarez-Gaume:2016vff}.
Also, the energy density of this parametrised solution for $p$ is, because of  \eqref{energydensity},
\begin{eqnarray}
\mathcal{H}&=&b_q
\sqrt{-p^{\prime}(x)^2+V(p(x)))}
\left(p^{\prime}(x)^2+2{V(p(x))}\right)\\
&=&\frac{T_{xx}}{2}+\frac{3}{2}b_q\sqrt{-p^{\prime}(x)^2+V(p(x)))}V(p)\\
&=&2{T_{xx}}-3b_q\sqrt{-p^{\prime}(x)^2+V(p(x)))}p^{\prime}(x)^2.
\end{eqnarray}
and the average energy density becomes, by using \eqref{baryonsin} and \eqref{baryoncos},
\begin{eqnarray}
\mathcal{E}=\frac{1}{\mathcal{V}}\int dx^i\, \mathcal{H}&=&
\frac{b_q\kappa^{3}}{2}+\frac{9}{16}(\rho_1\omega_1+\rho_2\omega_2)\\
&=&2b_q\kappa^3 -\frac{3b_q}{\mathcal{V}}\int dx^i\, \sqrt{-p^{\prime}(x)^2+V(p(x)))}p^{\prime}(x)^2\label{energy}
\end{eqnarray}

\subsection{Solving the equation of motion}
We restrict our attention to solutions for the $p$-field that have a point where $p^{\prime}(x)=0$.
This is because when we are
ultimately interested in infinite volume
as an approximation to finite volume at large charge, and
if we were to put the theory on $S^2$, we would have to impose the Neumann boundary condition for the $p$ field at some points.
Also, we can
use the translational invariance of the system in infinite volume, to set $p(0)=0$.

Now as we look for the lowest energy solution, in order to access the solution in the perturbative regime, we would like to set $\kappa$ and $\omega_1$ to be very close to $\omega_2$., i.e., we have two perturbative parameter $\epsilon$ and $\eta$, which are defined by
\begin{equation}
\epsilon=\frac{\omega_1}{\omega_2}-1,\quad
 \eta=\frac{\kappa}{\omega_2}-1.
\end{equation}
We take both of these parameters to be much less than $1$.

We will also take $\eta\muchlessthan\epsilon$, which is
equivalent to the condition $\r\ll 1 \muchlessthan \r\ll 2$.
This is \rwa{not} a necessary consistency condition for the solution to be in the regime of validity of the effective field theory; it is merely a condition to simplify the classical equation of motion sufficiently that we can verify easily that the ground state lies in the regime accessible to the EFT.  Indeed,
the EFT may be applicable for a larger range of $\r\ll 1 / \r\ll 2$, and we shall
comment later on this possibility.

As the ground state solution for $p$ is periodic, we only have to evaluate the amplitude of the derivative of $p$, hereafter called $v_0=p^{\prime}(0)$, and $p$ itself, hereafter called $p_0$, in spite of the difficulty of solving the full equation of motion analytically.
We assume that $p_0$ and $v_0$ are small, so that we can treat
them as perturbative deviations from the homogeneous solution, an assumption we will verify later.

Let us evaluate $p_0$ and $v_0$. When $p(x)=p_0$, the derivative of $p$ must be vanishing, and we have the algebraic equation for $p_0$,
\begin{equation}
\kappa^2=V(p_0)=\omega_2^2+(\omega_1^2-\omega_2^2)\sin^2(p_0)\iff
\sin(p_0)=\sqrt{\frac{(1+\eta)^2-1}{(1+\epsilon)^2-1}}.
\end{equation}
For small $p_0$, we have
\begin{equation}
p_0=\sqrt{\frac{(1+\eta)^2-1}{(1+\epsilon)^2-1}}(1+O(\eta/\epsilon))
\sim \sqrt{\frac{\eta}{\epsilon}}
\label{FirstEqRefMW}
\end{equation}
As for $v_0$, the maximal value for $p^{\prime}$ is achieved when $p=0$, so we have
\begin{equation}
-\frac{\kappa^6}{4}=(v_0^2-\omega_2^2)\left(v_0^2+\frac{\omega_2^2}{2}\right)^2.
\end{equation}
Solving the equation for $v_0$ which is small, we have
\begin{equation}
v_0={\omega_2}{\sqrt{(1+\eta)^6-1}}(1+O(\eta^2))\sim\sqrt{6\eta}\omega_2
\label{SecondEqRefMW}
\end{equation}
The spatial period of the solution, which is approximately $\ellz=p_0/v_0$ modulo multiplicative constants, becomes
\begin{equation}
\ellz\sim\frac{1}{\omega_2\sqrt{\epsilon}},
\end{equation}
which becomes infinite as $\epsilon$ goes to zero, \it  i.e., \rm we recover the homogeneous solution, as we must.

\subsection{Resolving the equation of motion at leading order}
We can also solve the equation of motion by noting
\eqref{FirstEqRefMW} and \eqref{SecondEqRefMW} and expanding all quantities to first order in $\eta$.
The equation of motion then becomes
\begin{equation}
2\eta=2\epsilon p(x)^2+\left(\frac{p^{\prime}(x)}{\omega_2}\right)^2,
\end{equation}
whose solution for $p$ is then
\begin{equation}
p(x)=\sqrt{\frac{\eta}{\epsilon}}\sin\left(\sqrt{2\epsilon}\omega_2 x\right).
\end{equation}

Using this to rewrite \eqref{baryonsin} and \eqref{baryonsin}, we have
\begin{eqnarray}
\rho_1&=&\frac{8b_q}{3\mathcal{V}}\omega_2^2(1+\epsilon)
\int dx^i\, \frac{\eta}{\epsilon}\left(1+\eta-2\eta\cos^2\left(
\sqrt{2\epsilon}\omega_2 x\right)\right)\sin^2\left(
\sqrt{2\epsilon}\omega_2 x\right)\\
&=&\frac{2b_q(1+\epsilon)}{3\epsilon}\eta(2-\eta)\omega_2^2\sim
\frac{4b_q\eta\omega_2^2}{3\epsilon}\left(1+\epsilon\right) \\
\rho_2&\sim&\frac{8b_q\omega_2^2}{3}-\frac{4b_q\eta\omega_2^2}{3\epsilon}
\end{eqnarray}
We have, as a consequence,
\begin{equation}
\omega_2=\sqrt{\frac{3(\rho_1+\rho_2)}{8b_q}}\left(1-\frac{\eta}{4}\right)
\end{equation}
and
\begin{equation}
\epsilon=\frac{\rho_1+\rho_2}{2\rho_1}\eta
\end{equation}
at first order in $\eta$.

\subsection{Minimization of energy}
We now check that the ground state configuration occurs away from the homogeneous solution.
By evaluating the average energy density for the solution of the equation of motion, we have
\begin{equation}
\mathcal{E}=b_q\kappa^3+\frac{b_q}{\mathcal{V}}
\int dx^i\, \sqrt{-p^{\prime}(x)^2+V(p(x))}(-p^{\prime}(x)^2+V(p(x)))
\end{equation}
Now the second term, using the equation of motion, becomes
\begin{eqnarray}
&&\frac{b_q}{\mathcal{V}}
\int dx^i\, \sqrt{-p^{\prime}(x)^2+V(p(x))}(-p^{\prime}(x)^2+V(p(x)))\\
&=&\frac{b_q\kappa^3}{\mathcal{V}}
\int dx^i\, \frac{1-p^{\prime}(x)^2/V(p(x))}{1+2p^{\prime}(x)^2/V(p(x))}
\end{eqnarray}
By using the fact that $|p^{\prime}(x)^2/V(p(x))|\sim \eta\muchlessthan 1$, we have, at first order in $\eta$,
\begin{eqnarray}
\mathcal{E}&=&b_q\kappa^3+\frac{b_q}{\mathcal{V}}
\int dx^i\, \kappa^3\left[1-\frac{3p^{\prime}(x)^2}{V(p(x))}\right]
=b_q\kappa^3\left[2-\frac{1}{\mathcal{V}}\int dx^i\, \frac{3p^{\prime}(x)^2}{V(p(x))}\right]\\
&\sim&b_q\omega_2^3(1+\eta)^3(2-3\eta)
\sim \frac{3\sqrt{3}}{8\sqrt{2b_q}}\left(\rho_1+\rho_2\right)^{3/2}\left(1+\frac{3}{4}\eta\right)
\end{eqnarray}
This means that the minimal value of the total energy is achieved at $\eta=0$.
The configuration associated with this minimiser at fixed charge densitied can be understood to have a constant amplitude $p_0=\frac{2\rho_1}{\rho_1+\rho_2}$ and an infinite spatial period, $\ellz\sim\frac{4\sqrt{b_q\rho_1}}{3(\rho_1+\rho_2)\sqrt{\eta}}\to\infty$.

The actual physical quantities which makes situation transparent is $\rho_1$, $\rho_2$, and $\ellz$,
so let us write $\eta$ and $\epsilon$ in terms of them and make the discussion above a bit clearer.
\begin{eqnarray}
p_0&=&\sqrt{\frac{\eta}{\epsilon}}=\sqrt{\frac{2\rho_1}{\rho_1+\rho_2}}\\
\eta&=&\frac{8b_q\rho_1}{3\ellz^2(\rho_1+\rho_2)^2}\\
\epsilon&=&\frac{4b_q}{3\ellz^2(\rho_1+\rho_2)}
\end{eqnarray}
You could also plug in these relations to the above argument for transparency.
Most importantly, $\mathcal{E}$ is given by
\begin{equation}
\mathcal{E}=\frac{3\sqrt{3}}{8\sqrt{2b_q}}\left(\rho_1+\rho_2\right)^{3/2}\left(1+\frac{3}{4}\frac{8b_q\rho_1}{3\ellz^2(\rho_1+\rho_2)^2}\right) + \cdots
\label{EnergyFormulav1}\end{equation}
so that it is apparent that $\ell$ as big as possible is the most favourable in terms of total energy.

In eq. \rr{EnergyFormulav1}, the dots $\cdots$ signify
omitted terms of order $\ell\uu{-3}$ and
smaller, and also terms of order ${{\r\ll 1\sqd}\over{(\r\ll 1 + \r\ll 2)\uu 3}}$ and
smaller in the limit where $\r\ll 1 \muchlessthan \r\ll 2$.
Relaxing this latter approximation may not
affect the result qualitatively.
The only relevant consideration is the range of $\r\ll 1 / \r\ll 2$ such
that the $\ell\uu{-2}$ term in \rr{EnergyFormulav1} has a positive coefficient.
The restriction to small $\r\ll 1 / \r\ll 2$ simply establishes that there is an
open set of values of $\r\ll 1 / \r\ll 2$ such that the true ground state
occurs at large $\ell$.

\section{Discussion}

We have investigated the ground state of the three-dimensional
critical $O(4)$ model in infinite volume, at general
charge densities $\r\ll{1,2}$.  To do this, we have used a large-charge 
effective theory described by a conformal sigma model, which is 
weakly coupled when describing observables on distance scales
large compared to $(\r\ll 1 + \r\ll 2)\uu{-\hh}$.  To understand
the true ground state, we have studied the helical solutions 
of the effective theory, \it i.e., \rm solutions preserving a combined
time translation and global symmetry rotation.
These solutions are equivalent to time-independent
solutions with general chemical potentials, and the true charged
ground state must always be among them.  For $\r\ll {1,2}$ both nonzero,
we find the ground state is always inhomogeneous.  In some
range of $\r\ll 1 / \r\ll 2$, the
inhomogeneity wishes to express itself on as large a distance scale
as possible; that is, there
is an energetic preference for arbitrarily large spatial period.

This outcome is a desirable one from the perspective of calculability.
The effective Lagrangian \rr{conformalsigma} is only
the first term in an infinite series of terms with higher
derivatives in the numerator, and powers of $|\pp q\dag\pp q|$ in
the denominator.  For helical solutions with spatial period
$\ell$, these corrections are suppressed by 
negative powers of $(\r\ll 1 + \r\ll 2) \ell\sqd$.  We see from eq. \rr{EnergyFormulav1} that the energetically favored solutions are those where 
the leading action \rr{conformalsigma} is most reliable. 

Ultimately, the main application of the large-charge effective
theory is to compute observables, such as
the ground state energy, in finite volume at
large but finite total charge.  The details will clearly
depend on the topology and geometry of the spatial
slice.  For toroidal spatial slices, the ground state
for sufficiently large global charges should be the
homogeneous solution whose spatial period is the
larger of the two cycles of the $T\uu 2$.  For a
spherical spatial slice, the situation is different, since
there are no isometries on $S\uu 2$ without fixed points, and so
there may be rapid variation of the fields near the fixed point of the isometry.
Nonetheless we expect the fields away from
the poles to have gradients set by the size of the sphere, rather than
the ultraviolet scale, 
and the dominant contribution to the ground state
energy on $S\uu 2$ to be calculable using the leading
order action \rr{conformalsigma}.

 Lastly, we emphasize again that we have only
shown this situation holds
for sufficiently small $\r\ll 1 / \r\ll 2$.  It would be good to find
the maximum possible range of ratios, for which long spatial periods are favored.

\section*{Acknowledgements}
The authors are grateful to Domenico Orlando and Susanne Reffert for valuable discussions.  
SM and MW acknowledge the support by JSPS Research Fellowship for Young Scientists.  The work of SH is supported by the World Premier
International Research Center Initiative (WPI Initiative), MEXT, Japan; by the JSPS Program for Advancing Strategic
International Networks to Accelerate the Circulation of Talented
Researchers;
and also
supported in part by JSPS KAKENHI Grant Numbers JP22740153, JP26400242. SH
is also grateful to the Physics Department at Harvard University, the Walter Burke Institute for Theoretical Physics at Caltech, and the Galileo Galilei Institute
for generous hospitality while this work was in progress.


\end{document}